\documentclass[a4]{article}
\usepackage[a4paper]{geometry}
\usepackage{amsfonts,amssymb}

\usepackage{cite}
\usepackage[cmex10]{amsmath}
\usepackage{amssymb}
\usepackage[pdftex]{graphicx}
\usepackage{array}
\usepackage[tight,footnotesize]{subfigure}
\usepackage{xcolor}

\def\grad{{\overrightarrow\nabla}}

\def\intersect{{\;\cap\;}}

\usepackage{alltt}

\def\impos{{\;\vcenter{\hbox{\rule{5mm}{0.2mm}}} \vcenter{\hbox{\rule{1.5mm}{1.5mm}}} \;}}

\def\lrarrow{\leftrightarrow \kern-8pt \rightarrow}

\def\2{\frac{1}{2}}

\def\Nav{\langle N\rangle_{\overline T}}

\def\beq{\begin{eqnarray}}
\def\eeq{\end{eqnarray}}
\def\2{\frac{1}{2}}

\def\lrarrow{\leftrightarrow \kern-8pt \rightarrow}

\def\frightarrow{\rightarrow \kern-11pt /~~}
\def\reducesto{\simeq \kern -3pt >}

\begin{document}
\newcommand{\strust}[1]{\stackrel{\tau:#1}{\longrightarrow}}
\newcommand{\trust}[1]{\stackrel{#1}{{\rm\bf ~Trusts~}}}
\newcommand{\promise}[1]{\xrightarrow{#1}}
\newcommand{\revpromise}[1]{\xleftarrow{#1} }
\newcommand{\assoc}[1]{{\xrightharpoondown{#1}} }
\newcommand{\rassoc}[1]{{\xleftharpoondown{#1}} }
\newcommand{\imposition}[1]{\stackrel{#1}{\impos}}
\newcommand{\scopepromise}[2]{\xrightarrow[#2]{#1}}
\newcommand{\handshake}[1]{\xleftrightarrow{#1} \kern-8pt \xrightarrow{} }
\newcommand{\cpromise}[1]{\stackrel{#1}{\frightarrow}}
\newcommand{\policy}{\stackrel{P}{\equiv}}
\newcommand{\field}[1]{\mathbf{#1}}
\newcommand{\bundle}[1]{\stackrel{#1}{\Longrightarrow}}
\def\Server{{\rm Agent}_1}
\def\Client{{\rm Seed}}
\def\Proxy{{\rm Agent}_2}
\def\Dispatcher{{\rm Dispatcher}}

\def\changes{} 

\title{A Quantitative Model Of Trust as a Predictor of Social Group Sizes and its Implications for Technology}
\date{\today}
\author{M. Burgess\\
ChiTek-i AS,\\
Oslo, Norway\\
~\\
and \\
~\\
R.I.M. Dunbar\\
Department of Experimental Psychology\\
University of Oxford\\
Radcliffe Quarter\\
Oxford OX2 6GG\\
UK}

\maketitle

\renewcommand{\arraystretch}{1.4}

\begin{abstract} \changes {The human capacity for working together and
    with tools builds on cognitive abilities that, while not unique to
    humans, are most developed in humans both in scale and plasticity.
    Our capacity to engage with collaborators and with technology
    requires a continuous expenditure of attentive work that we show
    may be understood in terms of what is heuristically argued as
    `trust' in socio-economic fields. By adopting a `social physics' of
    information approach, we are able to bring dimensional analysis to
    bear on an anthropological-economic issue. The cognitive-economic
    trade-off between group size and rate of attention to detail is
    the connection between these. This allows humans to scale
    cooperative effort across groups, from teams to communities, with
    a trade-off between group size and attention. We show here that an
    accurate concept of trust follows a bipartite `economy of work'
    model, and that this leads to correct predictions about the
    statistical distribution of group sizes in society.  Trust is
    essentially a cognitive-economic issue that depends on the memory
    cost of past behaviour and on the frequency of attentive policing
    of intent.  All this leads to the characteristic `fractal'
    structure for human communities. The balance between attraction to
    some alpha attractor and dispersion due to conflict fully explains
    data from all relevant sources. The implications of our method
    suggest a broad applicability beyond purely social groupings to
    general resource constrained interactions, e.g. in work,
    technology, cybernetics, and generalized socio-economic systems of
    all kinds.}

Keywords: trust, social physics of groups, scaling of dynamics, collaboration, social anthropology

\end{abstract}



\section{Introduction}

\changes{Economic ideas have proliferated over the past two hundred
  years in the modelling of all aspects of human pursuits. The
  principles of evolution have risen to a similar prominence, pointing the
  way in which dispersive environments select between alternatives
  without conventional agency. These two are not completely
  independent: the connection between them lies clearly in the
  statistical aggregations of chance interactions that wrestle (albeit
  blindly) for dominance. Evolutionary modelling makes use of economic
  ideas to discuss the mechanisms for selection of some `preferred'
  outcome, such as phenotype selection in biology\cite{dawkins1} by
  turning selection into a valuation problem. A population of agents
  may be thought of as a `market' superagent that selects
  product-offerings from a supply. However, in modern information
  theoretic terms, this arrangement forms a noisy information
  channel\cite{shannon1,cover1} that propagates successful `design
  semantics', regulating not only commercial diversity, but
  engineering choices and hence technological development in an
  evolutionary way too\cite{moneybook}. In a sense, technology is
  another kind of phenotype wanting for developmental selection, this
  time through the intermediary channel of human proxies. This is all
  explicit for genetic algorithms in software; we argue that the
  software of social interaction is wired into the same economics of
  cognition between agents.}

\changes{In their comprehensive review of the processes and
  evolutionary history of technological innovations, Koppl et al.
  \cite{koppl} remind us of two important features of technological
  evolution. One is that it involves a combinatorial process
  (characteristics compose into wholes). The second is that we should
  "include human actions among the ‘things’ being combined in the
  evolution of the technosphere." In other words, a complete theory of
  the things that make up the technosphere should include the human
  actions that form a crucial part of the process of technological
  evolution. Creation of knowledge is one part of understanding
  innovation, and dissemination (intentional) or diffusion (possibly
  unintentional) is another.}

\changes{The Achilles heel of economic arguments has always been the
  need to postulate qualitatively rational agents to explain
  behavioural coherence. Such rationality postulates conceal often
  unreasonably complex decision-making implications on the individual
  agent level in order to suppress the spectre of divergence of
  intent.  Unwanted subjectivity runs counter to the traditional
  objectivity of scientific method, and yet it is omnipresent at the
  scale of individual agents. One has to explain how scaling of
  specific intentions can produce a population dynamics that appear
  merely as a game of `economic' averages. Here we show how this
  objective universality derives from the episodic equilibration of
  information, over subjective contexts, for cognitive agent systems
  (whether biological or technological), without requiring any deeper
  explanation about individual characteristics.}

\changes{Group alignment of intent is the path by which science may
  try to legitimately argue for approximate statistical universality
  amongst semantic alternatives. Alignment may lead to compromise or
  partial consensus; in either case there is a mutual influence by
  information back-reaction.  One imagines that motivations are like
  arrows in the heads of humans, pointing in some abstract space, and
  that these interact through the dynamical process of socialization
  of criteria, combined with a de facto `agreement' of a majority of
  agents to align over what constitutes `economic value' or `fitness
  for purpose' in a cost-benefit sense. All this leads to a temporary
  coagulation of agents into groups with different effective
  intentions, i.e. functional alignments that define their behaviours
  in a granular manner. Neighbouring grains will further tend to align
  over the edges as group members diffuse from group to group as a
  result of contention in the group.  These variations have sometimes
  been modelled as magnetic vector domains in social physics by appeal
  to that dynamical analogy\cite{galam}. However, as the Chinese
  proverb has it: things collect into groups, while people divide into
  groups: humans also easily fall prey to contentious displays that
  eventually drive them out of groups. What emerges then is a process
  of group genesis and dispersion by `detailed balance' in which
  information is partially harmonized when agents come together, and
  that harmony is then disseminated when eventually they contend and
  move on to other groups, carrying some of that alignment with them.}

\changes{Our innovation here lies in {\em not} requiring agents to be
  motivated rational game players seeking to `win', but rather viewing
  their chance encounters like a gene pool ruled only by perseverance
  in the face of limited resources.  These are parsimonious
  assumptions\cite{durbin1,nei1,patriarca1}.  Over the past two
  decades, Promise Theory has resolved some of the issues over
  modelling of agent intent in technological and economic
  systems\cite{promisebook,moneybook}. It provides a formalization of
  agent modelling beyond simulations, placing dynamics and semantics
  on an analytical footing. To the causal philosopher, these matters
  might seem unduly technical, yet they have been major obstacles to
  scientific progress throughout the history of economic modelling.}

\changes{The question we have to ask is then: what makes one group of agents,
which align around a particular choice, thrive and another dwindle?
This is the key to understanding how outcomes will be favoured and
fulfill a niche in an ecosystem, whether biological or technological,
over time. This is the issue we address in this work. The
answer turns out to be not only beautifully simple, but cements a deep
connection between group dynamics and our capacity to relate to
technology.}

\changes{Economics has two main tools for dealing with selection: deterministic
Predator-Prey models and Game Theoretic models of strategy in the
tradition of Morgenstern and Von Neumann\cite{morgenstern1}. The
latter associates phenotypes with fixed payoff strategies. The
discovery of the Nash equilibrium\cite{nash1} made it possible to
partially escape from the naive assumption of rational agents that try to
maximize gain or minimize effort (see the works of
Hamilton\cite{hamilton1}, Axelrod\cite{axelrod1,axelrod2}, and
Maynard-Smith\cite{maynardsmith1} leading to evolutionarily stable
strategies). An initial bias towards cooperation as the answer to success was corrected by
pitting strategies against one another in tournaments. These revealed Anatol Rapport's tit-for-tat strategy
(an adversarial strategy) to be the winner\cite{axelrod1}. Today, we can use
hybrid stochastic models that go further still using Promise Theory to
underpin agent models without the need for player intelligence or
deterministic rules.}

\changes{In what follows, we show how evolutionary success of a process through
social group dynamics can be measured through the proxy of predictable
emergent group sizes. These are the analogue of Nash equilibria and
form a detailed balance for attachment-detachment that ultimately
derive on agent cognitive limits. Here cognitive needn't mean
specifically human mental abilities, only the ability for independent
agents to process information around them.  Our model stems from a
mixture of theoretical considerations, alongside empirical tests in a
wide range of fields in very different conditions, and also from a
speculative but compelling scaling argument concerning sampling rates
from neuroscience.  We use the autonomous agent framework of Promise
Theory to bring all these issues together into a universal
model, and thus eliminate the spectre of subjective bias in selection
criteria.}

\subsection{Innovation and human groups}

\changes{Although ideas may begin with creative reasoning either of
  individuals or conversations, most innovations are the product of
  teams who combine skills across different phases of the process. As
  Koppl et al. have noted, the penicillin revolution did not depend
  just on a lucky break by a careless technician in Fleming’s lab;
  indeed, several others had noted the same effect well before
  Fleming. It depended on both a novel insight by Fleming (namely
  that, if Penicillin killed bacteria on a plate, it might kill
  bacteria in a human body), then solving the more complex issue of
  purifying the active ingredient without destroying it (the work of
  Howard Florey’s lab a decade later) and finally finding a way to
  deliver it to patients. All this took place in bursts of episodic
  collaboration.}

\changes{Although both economists and other behavioural scientists are inclined
to view human activity as being a faceless, stochastic, infinite, and
even panmictic phenomenon, these assumptions are driven more by the
prerequisites for using the tools of differential calculus and common
algebra than by attention to reality.  Natural human populations are
in fact highly structured. More importantly, they are actually very
small in scale, and remain small scale even in the contemporary
post-industrial world \cite{dunbar6a}. The time available for
interaction and the cognitive limits on information-processing
capacity of the human brain impose extremely narrow limits on the
numbers of individuals with whom we interact, and hence know in the
sense of having meaningful, trust-based relationships.}

\changes{The result is that human networks have the structure of a
  series of concentric circles \cite{dunbar28}. Interaction
  frequencies decline as a steep negative exponential across the
  layers of the network \cite{dunbar10a}. When we learn something from
  someone else, that someone else is almost always a member of our
  network. This is likely to be so for two reasons: i) the more often
  we interact with someone, the more often we are likely to see or be
  told about a novel process or idea, and ii) because we trust
  familiar people more than unfamiliar ones, we are more likely to
  take notice of what they say. Our claim is that simplifying
  assumptions that ignore the individual are undoubtedly effective at
  the large scale of macro-economic processes, but at the
  micro-economic level they can be misleading.}

\changes{When the flow of information is person-to-person, as it is in the
processes of technology design and creation, then the internal
structuring of populations, organizations, work groups or even
personal social networks unavoidably creates eddies and slack waters
where innovations get trapped and not passed on. When this happens,
innovations, like genetic mutations, are likely to die out and go
extinct before they take hold in the population. A comparison of
innovation diffusion in panmictic and structured networks demonstrates
that, on average, new ideas are likely to take twice as many
generations to achieve fixation (i.e. full penetration in a network)
in structured networks compared to panmictic networks \cite{dunbar25}.}

\changes{Given that time is necessarily limited, this places constraints on the
sustainable size of communities, and hence the community's capacity to
invent and inform, even when structures like cities grow to enormous
populations\cite{cities,bettencourt2}. Understanding how these constraints limit rates
of change becomes important in developing a rigorous understanding of
the processes and possibilities of cultural and technological
evolution.}

\changes{This raises a number of important questions. How many people take part
in that process? Does the number involved make a difference to the
efficiency of task solution? Do too many cooks spoil the broth? Are
there constraints on this imposed by our evolutionary inheritance in
terms of cognitive abilities to manage interacting groups? 
Understanding how these constraints limit rates of change becomes
important in developing a rigorous understanding of the processes and
possibilities of cultural and technological evolution.}

\changes{Our analysis suggests two constraints on the processes of
technological innovation: i) Although involving large numbers of
people would greatly increase the speed and reach of innovation, in
fact the number than can be involved at any given time is severely
constrained; ii) The more people that are involved in a task, the more
likely it is that disputes will arise, slowing down the rate of
innovation. Progress comes through repeated determination and conflict
avoidance alongside often implicit cooperation, and is ultimately
based on a subtle dynamics of trust--which we define precisely.}

\subsection{Overview}

We describe a universal causal model for the dynamics of
abstract resource-limited agents, and show how it explains the now
well-documented link between social group sizes and their cognitive
abilities observed in humans and some other species\cite{dunbar0a,dunbar0}. Our
model has been tested in a large scale study of social interactions on
the Wikipedia social media platform\cite{burgessdunbar1}, and can be
compared to past studies of humans on large and small scales (for a
summary see \cite{dunbar3}).  The cumulative phenomenological evidence
for the existence of a hierarchy of human group sizes now covers a
broad range of scenarios; our model obtains an accurate fit for small and `large' $N$
compatible with relevant group sizes, based on a minimal set of general
assumptions.

While the preponderance of evidence to support invariant group sizes
arises from the arena of human structures, the implications of our
work probe more deeply into the economics of cognitive (or so-called
memory-feedback) processes at scale, so that the potential validity of
the argument goes far beyond human-to-human interaction, or a sense of
belonging, at various social group sizes: it concerns the scaling of
generalized human-technological systems\cite{treatise1,treatise2} and
the economics of interaction, belonging, and thus ownership on a
socio-economic level\cite{moneybook}.  Here we demonstrate the
underlying basis of constraints that act on human community size and
structure in ways that directly address the issues raised by Koppl et al
\cite{koppl}.  Understanding why human information exchange is
not, and cannot be, an ideal free mean field phenomenon has important
consequences for how we structure models in order to explore cultural
evolution either in a contemporary context or the historical context
of human evolution\cite{galam,sociophysics2}.  The integration of
Promise Theory is key to this understanding: it has been instrumental
in finding a proper dynamical scientific basis for a number of social
concepts, which have previously been addressed only in terms of
moral philosophy\cite{McNeilly1,atiyah1,owens1,sheinman1}, e.g.
intentionality, trust, authority, leadership, network growth, and
efficiency of communication.

By now, it's uncontroversial to say that the social dynamics of
all animals are deeply entwined with the neurological processes of
cognition that underpin them, and that this remains true for group
phenomena across multiple scales: from small to `large'.  A brain
plays a central (if sometimes implicit) role as both a calibrator and enabler of
`sticky' social behaviours as it provides a capacity for
distinguishing and recalling both other individuals and a more abstract
environmental capacity; it also acts as the cognitive glue that
keeps relationships alive, with memory keeping account of identity and
the trust that sustains attentive relationships. While it might be
tempting to promote the importance of animal relationships over other cognitive
challenges, we do not need to do so here. Although it's traditional
in the philosophy of the social sciences to focus on semantics of
moral determination including explanations of free will, etc, we
do not need to take that route either.  Our results apply to any
intentional activity that takes up some time-related capacity, whether
social or task related.

We make the case that group dynamics follow purely from the causal
interactions between agents with resource-limited processes, and that
the emergent group sizes fall into a statistical distribution by an
equilibrium principle of detailed balance\cite{reif1}.  Our model
reveals macroscopic group sizes to be an emergent outcome of agent
{\em memory processes}, in an information physics
sense\cite{grimmett1}. The causal behaviour enabled by
memory, shapes group distributions through an effective
attraction/repulsion between agents is induced by an `economic accounting
history' own efforts, summarized as an accounting potential, which is
related to their kinetics---a measure of (in)attentiveness \cite{dunbar1}.  The
effects of this energy-like accounting parameter mirror the semantics
of `trust' between the agents.  The result may be used to
calculate a distribution for the main features of group
dynamics: the hierarchy of group sizes, for different levels of
relative attentiveness.

\begin{itemize}
\item We begin by
summarizing some relevant phenomenology of group dynamics that
provides an observational basis for our model. 
\item We then sketch out the
axioms of Promise Theory as a language for the description of agents
and their intentional process alignments with activities as a framework
for representing the link between the dynamics and semantics of processes.
The formal language of promises give us both an interpretation and
representation of intentionality across agents of any scale, without
the need for spurious interpretations of free will.  

\item We then describe how
agents form groups by binding into network structures with different
topologies. We shall claim that group phenomenology is dominated by a topology of
simple star networks. This is somewhat unexpected from a network science perspective, 
but it makes sense in the context of Promise Theory, as agents join
up by seeding of intent. 

\item Finally, we show how straightforward dimensional analysis constrains the scaling of
process rates for attachment and detachment to and from
groups. From the dimensions of process rates, an effective energy
parameter emerges in potential and kinetic forms to play the role of {\em trust} in its two forms
(trustworthiness meaning reliability and kinetic trust meaning inattentiveness) that quantifies rates.  
\end{itemize}
The
average of a detailed balance between attachment and detachment gives
a scale-free expression for the scaling of group sizes in good
agreement with data from Wikipedia studies, and in
agreement with evidence from other group studies and from the
neuroscience of attention.

\section{Phenomenology}

Data about social  group processes and their size distributions stem from a
number of sources, each lending independent support to
the hypothesis that animals self-organize into
hierarchies of social groups at very specific scales \cite{dunbar6a,dunbar13a,dunbar14a}.
We summarize some key background points below.

\subsection{Groups and their scaling}

An important starting point for understanding cultural transmission is
the size and structure of human communities. Conventionally, most
evolutionary and economic approaches assume panmictic social
arrangements, usually formulated as mean field models. However,
natural human groups (as defined by the number of individuals with
whom someone has meaningful relationships) are in fact very small---of
the order of 150 people \cite{dunbar6a}. This value forms part of a general
pattern in primates in which a species' typical social group size
correlates with the size of its brain (more specifically, its
neocortex) \cite{dunbar0}, a relationship known as the social brain
hypothesis. This relationship reflects the cognitive demands imposed
by the need to manage the relationships involved in groups of
different size \cite{jerison,byrne1}.  

A second dimension to the social brain hypothesis is that it has a
fractal structure. Not only does the distribution of primate social
group sizes follow a `fractal' distribution, but these groups are themselves
`fractally' structured internally \cite{dunbar18}. In the case of human
communities, this is manifested as a series of hierarchically
inclusive layers at 5, 15, 50 and 150 \cite{dunbar6a,dunbar14a}. In
humans and primates, these layers emerge out of the differential
frequencies with which individuals interact with other members of their
group \cite{dunbar20,dunbar17}. 

The relationships that characterize
the different layers differ not only in contact frequency but also in
emotional closeness, trust, and willingness to act altruistically
\cite{dunbar20}. Relationships of different emotional quality and
locus within an individual's network are processed in different
components of the main neural network that manages social
relationships (the default mode neural
network)\cite{dunbar21,dunbar19}.
Thus, this fractal structure directly influences the rate at which
information as well as emotional engagement flows through a social
community. If we wish to understand the factors that influence
the creative processes that produce new technology and the processes
whereby knowledge of these innovations spread through populations, we
need to understand why and how human communities are structured, and
how this structure influences cultural transmission. Here, we focus on
the first part of this programme.

To study these processes at different scales, we use the
physics of scale analysis\cite{scale2}.  Scaling relations depend
mostly on dimensionless ratios of measurable quantities (see also the
discussion by West \cite{gwest1,westscale}), because different
engineering dimensions (mass, length, time, etc) play different roles
and thus imply altered meanings. We therefore seek relevant counting
parameters that enable or limit growth.

\subsection{Economics of trust and work}

\changes{The economic argument for group behaviours relies on the principle of
aversion to work cost, i.e. statistical cost minimization, where cost can be measured in work. Work is well
defined in physics and economic theory was modelled firmly on those
definitions\cite{mirowski}.}  Animals including humans sustain
relationships with one another through the kinetics of activities like
grooming, talking, working together, and so on. We refer to all of
these as generalized grooming. This involves an expenditure of work,
in a physics sense.  When mutually beneficial, we understand that such
activities appear to build trust between them--hence there is a direct
association between trust and energy.  Like most humanistic notions,
the history of trust has been dominated by ideas of moral philosophy
\cite{fukuyamatrust,putnamtrust,wendystone,WhatisSocialCapitalAComprehensiveReviewoftheConcept}.
Some progress has been made in social sciences by attempting to model
certain scenarios by analogy to simple physical
systems\cite{galam,sociophysics2}. However, a more agent-centric view
of trust can be given by using Promise Theory to capture the simple
information relationship between trust and
intentionality\cite{trustnotes}. In this view, the trust about some
subject $X$ is related to the work saved by not verifying $X$
\cite{trustnotes}. However complex the semantics of these processes,
they ultimately flatten out insofar as they simply involve different
expenditures of effort. This is why complex behaviours can ultimately
be reduced to simple numbers like effort or group size, allowing us to
measure them.

In cognitive terms, the accounting of relationships involves
recognition of identities and thus memory in order to distinguish
agents and their intentions.  Thus, on the micro-scale, processes are `memory processes', i.e.
not of the basic Markov type\cite{grimmett1}. They
require a history of past interactions using both internal (neural)
and external (stigmergic) memory, and are thus resource-constrained by cognitive ability.

\changes{Promise Theory goes beyond economic contract
  theory\cite{contracts,fried1} by seeking to represent intentional
  semantics of agents with cost in a manner compatible with the tenets
  of Information Theory\cite{shannon1,cover1}, by replacing
  conventional (deterministic) differential equations of economics
  and rational minima of equilibrium games with something the takes
  account of both for finite resolution relationships.  Much of
  economic modelling originated in the naive reading of physics by
  analogy\cite{mirowski} in an effort to replicate its successes;
  this, in turn, builds on memoryless Markov processes for their
  simple universality. However, a true information model requires
  that agents in a relationship be engaged in a series of on-going
  interactions with intentional alignments, in which each one samples the other's behavioural
  states in order to assess alignment with promise-keeping trustworthiness.  This
  cyclic sampling has a rate of work that associates trustworthiness
  with an energy expenditure.  The attention rate or agent sampling
  rate relates to the resolution of the information by the Nyquist
  Theorem. Furthermore, it relates to cognitive energy expenditure for
  agents, as acknowledged in neuroscience\cite{mitchell1}.}

\changes{As we show below, trust-related work is the first cost-benefit parameter involved in agent dynamics.
Time is related to work on a number of levels. There are two main
timescales at which system state relates to evolutionary change: i)
the individual cognition of agent to agent interactions that are
related to brain oscillations in humans, and ii) the impact of
environmental pressures that help to define and shape the intentional
behaviour of agents. This is universal and independent of group size.}

\changes{The economics of social group size also enter through discrete
counting scale parameters. Ultimately these are limited by the power output of an
agent's cognitive effort in sustaining the level of attention required to
stabilize coherent activity between agents working in the same group.  The
collective benefits of grouping for individuals thus influence group
dynamics in different ways.  For herd animals, flocking together
results in a boundary formation between group and exterior that
potentially protects the members within the boundary. The presence of
an alpha leader not only forms a seed for gathering around but also
offers small-group protection and may limit conflicts and in-fighting,
providing stability like a memory function.  Advanced agent groups may
also be able to use cooperation by delegation of coordinated tasks and
capabilities to achieve a goal not possible for an individual.  This
is the argument referred to above for bringing innovation to market in
the case of human cooperation.  The combinatoric strategies for agent
delegation and cooperation in software systems and artificial
intelligence are also well known.  In all cases, the larger the group
size for agents the more work effort is needed to coordinate and
maintain functional stability.  Thus the economics of agent
cost-benefit optimization is bound to be a non-linear function of
agent number.  What is remarkable is that, in the case of
trust-contingent economics, this is has a universal character for
agents that are approximately similar in their interaction rates---but
not in the way authors conventionally discuss trust.}

\subsection{Group size $N$ as a proxy for dynamics over different timescales}

We don't need to know precisely how a brain (or other central control structure
in the corresponding cognitive role) might be constrained to
deal with a certain number of relationships, only that there is some
finite capacity limit. This is because what matters is the relative {\em rate} of
work---whatever the process may be.  The evidence from the Wikipedia study\cite{burgessdunbar1}
suggests that the cost of interacting can become too high due to
{\em contention} between agents, once they have grouped, and in that case we use the
semantics of maximal contention to anchor the controlling
scale, represented as $\Nav$ in equation (\ref{formula}) below. However, in other cases there
may be some differences in the reason for group break up. 
Thus, there is some
freedom to define the semantics of a controlling parameter scale by
absorbing interpretations into a dimensionless parameter $\beta$ which
plays the role of a fixed fraction of the work. As long as we express
relative measurements in dimensionless terms with universal meanings,
we can eliminate dependencies like this by computing only relative
quantities so that such details cancel out. Our approach is therefore
to convert effective work/effort ratios into effective group number
ratios. As long as quantities are scale invariant, relative answers
are then expected to be independent of details such as species,
capability etc, up to some dimensionless corrective factor---at least
insofar as we assume that the cognitive processes are based on the
same scaling principles across species.

Time is a parameter in any dynamical system and total work accumulates linearly
with the frequency of interactions multiplied by the number of agents
in a group, so the amount of work affordable by any individual,
interleaved between agents over a particular time interval, scales
inversely with the size of the group for that process and the invested
time cost of each interaction, 
\beq
N \propto \text{Work} \times {\Delta t} 
\eeq
or in terms of interaction frequency (level of intimacy) $f \sim 1/\Delta t$ into a dimensionless form:
\beq
\frac{N}{N_0} \propto \frac{\text{Work}}{W_0} \times \frac{f_0}{f},
\eeq
i.e.  the fewer
members we lavish attention on, the more time we have to do so and
with greater effort\cite{dunbar11a,dunbar12a}.  Groups may thus be ephemeral or long lived, but
we can separate and scale these accordingly by making the connection between
trust and the investment of time and effort. In this way, we use an old near-equilibrium
scaling argument to transform a short-term non-equilibrium system into an effective long-term equilibrium
for an adaptive agent system (a brain) that can adjust its relative interaction sampling rate
over short times. The effective long term behaviour thus becomes a Boltzmann statistical mechanics problem.
Conversational clusters,
for example \cite{dunbar7a}, involve processes other than long
term friendships or the persistent associations with kin, tribe, or
work associates, etc \cite{burgessdunbar1}. 
For each scale, there will be a similar relationship with different
proportionalities.  Dunbar proposed, in this way, that an equilibrium group size $N$
could be taken as a time-independent proxy for the complex
time-dependent cognitive and social processes at work in social
dynamics \cite{dunbar1}.  This has been demonstrated at
length in the literature, and for humans one finds an average base number $N$ for
attentive human groups lying somewhere between $N=4$ and $N=5$
\cite{dunbar15a,dunbar4a,dunbar8a,dunbar9a,dunbar16a}.  This may be
compared with the computed outcome of our model in section
\ref{maxcurve}.  The limit on conversational group size, for instance, appears to be
set directly by the capacity to manage the mental states or viewpoints
of other individuals \cite{dunbar8a}.  From groups involved in
Wikipedia-editing, familiar group patterns for humans were observed with
scaling close to that for conversational dynamics\cite{burgessdunbar1}.

Notice that, in all cases, trust is built through the social process,
which effectively leads to a repetitive training of memory, by the
rehearsal of the social bonds, forging deeper links between process
and brain activity.  Memory may take the form of internal recollections, or
stigmergic traces left by society that transmit influence
from independent episode to independent episode on a group level: e.g.
the building of a department of urban planning transmits a certain
behavioural norm from episodic generation to generation. This is well understood
from Swarm Intelligence and Axelrod's evolutionary game economics\cite{axelrod3,siriAIMS1}.
Interpreting positive or negative social encounters may involve
complex semantics, particularly for humans, but for the purpose of
determining group size, all that matters is time spent on $X$ (for
some intentional behaviour $X$), and we argue once again that the
details flatten out into an effective summarial currency of `trust' as
this increasing familiarity stabilizes over many cases
\cite{dunbar10a,burgesstrust}. This is what data scientists refer to
as a dimensional reduction. We can thus talk about trust specifically
for any distinguishable process of keeping some promise of $X$. It
measures the mutual alignment of agents on the subject of $X$.
Promises effectively act like a spanning set of coordinate axes over a
space of intentionality, factoring out the complexities of semantic
interpretation.  The quantitative interpretation of trust as an
effective energy/work parameter is now natural, since energy is a
complementary variable to temporality in physics and tracks the local
spending of work over a given time.

\subsection{The Wikipedia study}

\changes{Since the group scaling hypothesis was proposed by Dunbar,
  many studies, from analyses of conversations to communal groups,
  have broadly confirmed the facts (e.g. see the summary in
  \cite{dunbar3}).  As is the norm for sociological studies, the
  numbers of participants was typically limited both by the idea
  itself and by availability of willing participants, thus limiting
  the accuracy and universality of the group size results. It became
  possible to surpass this limitation thanks to the fortuitous
  discovery of an independent study concerning the role of trust in
  unrelated work by Burgess using Promise Theory\cite{trustnotes}.
  Since the role of trust online opens the door to large data sources
  that are fully randomized by global service access, the link changed
  the nature of experimental evidence substantially. In the online
  world, one can study ad hoc communities that are completely unbiased
  by geography of social class. Using data from Wikipedia change
  logs\cite{NLnet}, Burgess documented the behaviour of users in
  relation to their activity levels and signs of contention and found
  an unexpected group behaviour that was tantalizingly close to the
  Dunbar numbers\cite{dunbar6a}. The results were summarized in \cite{burgessdunbar1}
  so we shall not repeat the details here.  Rather, we summarize
  their significance.}

\changes{Initially, one might imagine from conventional views on trust
  that users of Wikipedia come together to help one another work on a
  page of information, and that they do so because they trust one
  another.  This picture turns out to be completely upside down, being
  rooted in conventional platitudes that characterize the literature
  of trust. What the data showed was rather than someone starts a page
  on Wikipedia ad hoc, usually alone.  After some time, another user
  will notice the page and be attracted to come and contest or alter
  what was written. The network structure is the simplest one: a seed
  or alpha leader attracts followers to join, not friends of friends
  etc, but random attraction in the manner of a lighthouse attractor.
  The activity triggers others to come, including editors on the site,
  and a burst of arguments and conflicts begins.  The group grows
  until it reaches a certain size, and then people begin to
  leave---perhaps weary of the work involved in defending their
  choices. The number of negative comments and
  undo-operations may be interpreted as evidence of mistrust.  Eventually the group
  disbands, some time passes and then it starts over again.  Work is
  not continuous; it is quite episodic. The result of this interaction
  leaves all of the users closer to a common view than before, whether
  willingly or unwillingly because the product of their efforts is
  shared and could easily be undone. Survival favours consensus.}

\changes{One sees the same basic behaviour in all topics, not just
  controversial ones: in everything from mathematics to pop
  celebrities, pages attract users who come and stay for a while
  because they are mistrusting of one another.  Plotting the
  distribution of users who are active in each episode, one obtains a
  distribution shown in figure \ref{datafit}. The picture indicates
  that users are drawn to the perceived potential of the Wikipedia
  platform, but what makes them keep coming back is their mistrust of
  other users. We explain this in detail below. What's clear is that,
  if users merely trusted one another, they wouldn't need to
  continually check each others changes and contest them, they could
  simply go off and do something else knowing that everything would
  turn out well. Instead, the persistent lifetime of the repeated
  refereeing contests amounts to a kinetic work of attentiveness, and
  applying the theory below we are able to predict the distribution of
  users per episode with unrivalled accuracy, thanks to data sets of
  hundreds of thousands of individual users.  This is
  unprecedented access for a social study.}.

\section{Theory}

We can now describe the theoretical model that predicts group scaling
in accordance with the phenomenology.  We wish to abstract away as
many non-pertinent details as possible in order to discuss the scaling
of group formation for maximal universality.

As we collect the dimensionless quantities of autonomous (causally independent)
agents into a single statistical ensemble, we effectively project individual
contributions into a space
of common relative characteristics. Individualities are dispersed across
a probabilistic distribution of alignments for each `promised' property.
We are not able to know much about these effective promises, but this
is of no consequence since we only need the rates at which the promises are
kept. This gives a representation of the work done, and which may be
summarized into a parameter that we shall simply call `trust'.
In this way, we express the large scale statistical essence of group
formation as a physics of processes rather than a taxonomy of animal
attributes. We treat individuals simply as abstract agents and we turn
to Promise Theory as a suitable description of agents and
scaling\cite{burgessdsom2005,promisebook}.

\subsection{Process coupling strength and pair bonding strength}

In physics, the principal of separation of scales refers to the phenomenon whereby
sufficiently weak couplings between active agents in a dynamical
systems leads to qualitatively different behaviours on small and large
scales. This is formalized by the renormalization group and
dimensional analysis\cite{wilson1,scale2}.
In our social dynamics, the effective coupling strength refers to the
process of reinforcing a social bond with a small or large relative
interaction frequency (which we may associate with the semantics of a
relative intimacy), so the definition of a coupling strength amounts
to an interaction rate or an energy/trust scale.

The related principle of dynamical similitude observes that similar behaviours
have similar explanations. Similarity here is gauged by
the identification of scale-free or dimensionless variables\cite{scale2}.
The scaling of agency, from individual to group, implies that collective agents may behave effectively as
single agents on a new scale, as long as the structure of their
dynamics is similar. Dimensionless variables control these
universal features.

Mentally taxing relationships are thus related to more intimate
relationships, because they both depend on the same finite internal
agent resources. This is probably why contentious interactions form a
natural scale for group formation in Wikipedia\cite{burgessdunbar1}.
Moreover, because the work involved in agent processes cares nothing for whether a
relationship is physical or abstract, an agent's attachment to an
abstract goal, such as a task or a tool may be essentially similar to
its relationship to another agent in terms of work and trust,
depending on the agent's capacity for internal representation of its
semantic world.

\subsection{Bottom up causation}

In order to understand the level of determinism in our main result,
equation (\ref{formula}), a brief note about causality is in order.
The long standing tradition in Natural Science is to
assume the concept of a `force' or `command message' as the mediator
of a {\em necessary influence} in the dynamics of agents. More recently, there has
been a significant shift away from this deontic view of top down
causation to one of bottom up `emergence'.  This is not a matter of
taste, but rather because consistency with the principles of local
autonomy requires it\cite{promisebook}.

A Promise Theory model incorporates this bottom up perspective in its
axioms. Thus agent determinism is from the inside out rather than the
outside in---what one could call `voluntary cooperation' in human
parlance. It also means that promise theoretic models are directly
compatible with related descriptions in Game Theory\cite{myerson1},
Graph Theory\cite{berge1}, Network Science\cite{albert1}, and
Information Theory\cite{shannon1}, by contrast with Modal
Logics\cite{modallogic}.

\subsection{Alignment of intent and promises}

In Promise Theory, intentionality is represented by formalization of
stylized `promises' as a representation of intent. These express a
`direction' of intent, from one agent to others, and are defined
relative to a space of possible outcomes for the process concerned
with keeping the promises.

Any agent $A_1$ may promise some $X_1$ to another agent $A_2$ freely and independently.
$A_2$ determines freely and independently whether or not to accept this. 
Promises from one agent to another are called offers or donor promises and are denoted with a (+)
sign. Promises to accept another's offer are called acceptances or receptor promises and are denoted with
a (-) sign. Promises are local, i.e. each agent's promises can be kept by processes of the agent making the promise (see point 4).
Suppose it promises
that it will accept $X_2$, then we write:
\beq
A_1 &\promise{+X_1}& A_2\\
A_2 &\promise{-X_2}& A_1.
\eeq
Together these two promises are the necessary condition for a binding causal interaction.
Autonomy implies that the agents may not be completely aligned.
Assuming the agent $A_2$ accepts some amount (denoted $-X_2$) of what is offered (denoted $+X_1$), then
their binding has strength $X_1\intersect X_2$. This is the mutually intended but {\em unidirectional} outcome
for agents in a relationship. 
The main assumptions of Promise Theory can be summarized as follows:
\begin{enumerate}
\item {\em Agents}: every active player is an agent. Agents are autonomous, or causally independent of one another.
Agents have internal resources to form intentions and execute influence on other agents.

\item {\em Intent and promises}: an agent's intention, which is made public to a select group of other agents, is called a promise.
Promises are directed to one or more other agents and thus form a directed graph of intent.

\item {\em Assessment}: each agent forms its own assessment of whether
  some promise is kept or not, meaning agents are not necessarily in
  perfect alignment about their understanding of what occurred. Agents
  may or may not be faithful judges of promised information.
  Assessments are potentially as complex as the agents that make them.
  They involve processes of individual judgement, use of reputation,
  local costs and so on. The autonomy of agents speaks against the
  simplistic logic of exactly $X$, true or false.

\item {\em Strong autonomy}: no agent can make a promise on behalf of
any other. Promises thus only affect an agent's own state. Agents are not obliged to accept one another's promises. Thus agents
maintain an extreme form of the principle of locality of action.

\end{enumerate}
We make use of these points implicitly in assessing and counting the alignment
of agents in what follows.  

It remains for a statistical
model to consistently sum interactions to yield bulk results.  
For our model of trust, we note that the assessment of trustworthiness is updated
when agents are assessed to keep their promises to a high enough degree to motivate
repeated encounters. After binding, residual mistrust translates into an ongoing kinetic process
of investing work in checking the promise outcomes repeatedly, as an ongoing learning process.
Monitoring relationships becomes effectively a tax on cooperation.
In this way, the statistical
properties of Promise Theory become a kind of information physics, where
agents bind together virtually by the promise bindings that
persist over time.
The outcome requires the attention of both parties, with donor and receptor promises, 
and involves time as an implicit resource.

\subsection{Trust as a dynamical action-attention potential}

Given bindings, dimensional analysis provides the framework establishing common semantics
for causally independent agents, each with their own systems of assessment.
All assessments of change can be decomposed to
combinations of a few basic properties regarded as `innate' to physics, namely
mass, length, time, etc. The role of time is the most important for
group formation, principally because it is closely associated with
work done.  The counting of any repeating process over time has to
follow this universal dimensional analysis.

In physics, the relationship between stored potential and kinetic
energy is basically a dimensional equivalence between what work is
accumulated over time to be reused as a potential, and how previously
saved potential spent over a shorter timescale as a `kinetic'
activity.
In Newton's classical continuum language of moving bodies, a force $F$
applied over a path length $dx$ in some parameter space is equivalent
to a directional impulse $dp$. If one assumes a process rate or velocity $\vec
v=\vec{dx}/dt$, where $\vec{dx}$ represents the direction of an
intention in the space of outcomes, this settles the accounting of the
quantities with respect to time.  The usual `Newtonian' conventions
follow from the observation that a change in potential energy
(defining a force) has the same dimensions as a change in kinetic
energy. The equivalence can be seen in a number of steps that are
exact in the continuum limit (see equation \ref{newton} below).  For trust, we can use identical symbols
with only a renaming: $V$ is a trustworthiness of trust potential, and
$\overline T$ is the attentiveness or kinetic trust.

\beq
dV = \grad V\cdot\vec dx = \vec F\cdot \vec{dx} &=& \vec F\cdot \vec{v}\,dt\nonumber\\
&=& \frac{\vec{dp}}{dt}\cdot \vec v\;dt\nonumber\\
&=& \vec v\cdot \vec{dp}\nonumber\\
&=& m\vec v\cdot \vec{dv}\nonumber\\
&=& \2m d(\vec v\cdot \vec v)\nonumber\\
&=& d\left(\2 mv^2\right)\nonumber\\
&=& d\overline T.  \label{newton}
\eeq
In Newtonian language, a stored potential $V$ is the amount of
currency accumulated, equivalent to the action of some influence $F$
towards the execution of a directed process along $x$.  The latter
only defines $F$ in mechanics, and is not important here\footnote{The
  equivalence is the basis of Newton's second law and can be viewed as a
  definition of a concept of influence as a force as $F=m\,dv/dt$, for
  a proportionality constant mass $m$. The meaning of influence in
  modern physics is somewhat more complicated than this, but the
  essence of a force is contained in this dimensional
  equivalence\cite{jammer2}.  }. On the other hand, the speed at which
this is executed gives a velocity $v$, and its intrinsic inertia
represented by $m$, which becomes a constant of proportionality.

By reinterpreting the meaning of the quantities, we can use this dimensional equivalence to relate
trustworthiness $V$ to kinetic attention rate (mistrust) $\overline T$. The latter
is the rate at which an agent expends work to check whether
the promised intentions are in line with expectations or not.

Our ability to capture information about a process, distributed over a
population of very different contexts and agents, depends on being
able to reduce it to the counting of simple scales that can be shared
between all parts of the system.  Adopting this formal approach to
quantifying trust makes this possible.  It also provides a causal
explanation for why for the study of social physics by analogy has
some success \cite{galam,sociophysics2}.  We use this to overlay the
effects of every possible pair of agents engaged in an extended
interaction, each with its own effective clock and measures, onto a
single statistical process with a single common clock and measure, by
assuming that the semantics of of mass, length, and time are common to
all processes so that measurements calibrated to individual standards
can all be combined meaningfully.

\section{Agent model of group alignment}

\begin{figure}[ht]
\begin{center}
\includegraphics[width=12cm]{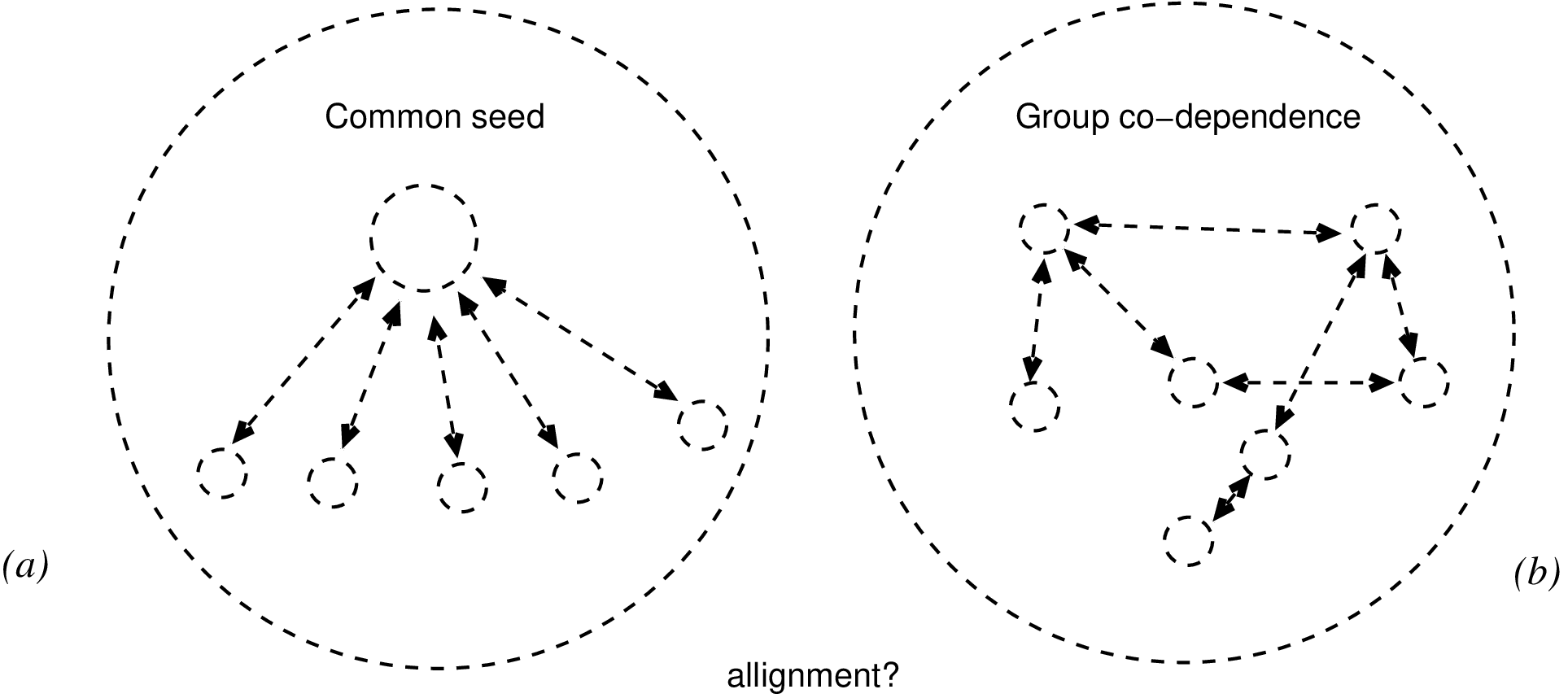}
\caption{\small Groups form either because agents come together
  independently attracted to contribute to a common cause (like
  fighting a common enemy or working on a common product), or they
  form emergent clusters by pairwise percolation of promise
  relationships. The alignment of intent between autonomous agents is
  co-dependent with their interactions.  This applies to agents of any
  type, whether biological, sociological, or cybernetic
  (human-technological). In our model, we assume the left hand picture
  of attraction in which `mistrust' of the central `seed' promise
  drives increased attention and potentially proximity as a secondary
  effect.\label{coop}}
\end{center}
\end{figure}

We can now build the elements in the foregoing sections into a
method for counting network bonds in a bulk population.  Figure
\ref{coop} shows two ways in which groups could form from individual
agents.
A group that follows a single
leader or interacts one at a time is different from a group that tries
to maintain global coherence all at once and all the time. The former
is quite rigid and the latter redundant but expensive. For $N$ agents, the cost ranges from order $N$ to $N^2$.
By going through an intermediary, the work cost to agents in coming
together around a single seed is minimized to $O(N)$, where $N$ is the
number of agents in the emergent group. If agents had to maintain
contact with every other in a group, it would cost them up to
$O(N(N-1))$, which is significantly greater. Thus leaders and hubs,
even abstract totems, serve an economic purpose in group coherence.
The latter form of group cohesion is quadratically expensive to
sustain beyond $N=3$, as the cost of predictable assurances
rises as $N^2$.

In our discussion, we find the loose hierarchical association of the
first case to be the cost that provides the best agreement with data,
as well as being the simplest causal mechanism for attachment. This linearity
may also explain why the `fractal' scaling series can be sustained: a
quadratic pattern could not be scale invariant in this way.

\subsection{Calibration and consensus through common dependency}

\begin{figure}[ht]
\begin{center}
\includegraphics[width=8cm]{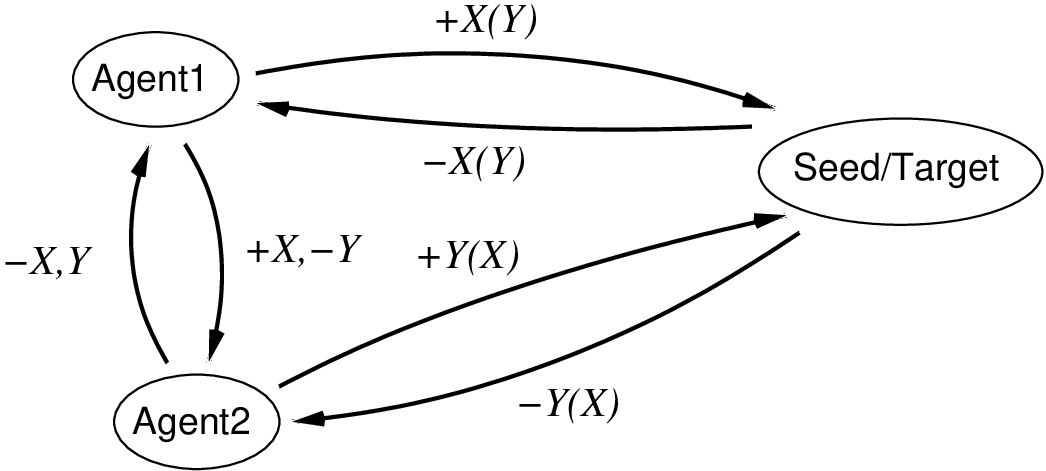}
\caption{\small The cheapest approach to alignment over a group can be seeded 
by a basic common calibration triangle in Promise Theory
allows two agents to work together on behalf of a third, or allows a third to act
as a seed effectively bringing them into alignment $X=Y$.
From Promise Theory, one would expect opportunistic dyadic structures 
$N=2$ for compositional or symbiotic specialization, 
with more important coordinated structures built from equilibrated/cross-checked
triads $N=3$.
\label{inter2}}
\end{center}
\end{figure}

Consider the primitive pattern involving three agents shown in figure
\ref{inter2}. The triangle of promises is the maximum
coordination for an instance of three agents. This is the configuration by which they
can maintain consistent information and claim to `agree' with one
another. It is called the Law of Conditional Assistance in Promise
Theory. It represents a configuration of voluntary cooperation respecting the
autonomy of the agents.  $\Server$ promises an intended
outcome $X$, based on the other agent's intent to supply $Y$' in the
most general sense.  The intended outcome $X$ could involve watching
over the group, performing some work on its behalf, etc. Essentially,
it requires paying attention to the promise and allocating time
resources.  $\Server$ also promises to make use of the promise $Y$
provided by $\Client$, which could simply be access to its personal
space, or the ability to perform some service for it.  We can use the
shorthand notation for the directed promises: \beq \left.
\begin{array}{c}
\pi_X: ~ \Server \promise{+X|Y} \Client\\
\pi_Y: ~ \Server \promise{-Y} \Client
\end{array}
\right\rbrace \equiv \Server \promise{+X(Y)} \Client.  
\eeq 
to represent the conditional promise of $X$ given $Y$, together with the promise to accept $Y$
if offered. In other words, `I will keep the promise of $X$ with the assistance of
another, who in turn helps me by supplying $Y$, written $+X|Y$, and I promise you that I am
accepting such help $-Y$'. The full collaboration now takes the form\cite{promisebook}:
\beq
\Server &\promise{+X(Y)}& \Client \label{pull1}\nonumber\\  
\Server &\promise{-Y, +X}& \Proxy \label{pull4}\nonumber\\
\Proxy  &\promise{+Y,-X}& \Server \label{pull3}\nonumber\\
\Proxy &\promise{+Y(X)}& \Client \label{pull5}\nonumber\\
\Client &\promise{-X(Y)}& \Server \label{pull2}\nonumber\\
\Client &\promise{-Y(X)}& \Proxy\label{pull6}
\eeq
Notice the symmetries between $\pm$ in the promise collaboration of
equilibrium state, and between $X,Y$ indicating the complementarity
of the promises. 
The maximal cost of this configuration is close to the square of the number of agents.
Such a cost is unsustainable for large numbers.

We underline that these promise triangles are not related to earlier concepts
of significant threes in social science.  Simmel introduced a notion of triads in social
systems\cite{simmelsoc} as a speculative pattern describing semantic
complementarity. Another triad theory of sentiment relations in social
balance theory was proposed in
\cite{fritz1,fritz2,khanafiah2004social} as a rule of threes proposed
for the social sciences\cite{triads1}. 
Triadic agent molecules have
been proposed many times as basic control units for social networks,
and have been implicated in group formation.
The triangle relation we use here effectively promises the semantics of an
{\em equivalence relation} between pairs of agents. The promise triangle
is a form of covalent bond, in chemistry parlance: two agents are held
together by mutual bonds with an intermediate third party, which calibrates their involvement.

Agents are most simply glued together by mutual association with a
proxy or go-between (this is a covalent bond, in chemistry parlance). The intermediary might
be a physical agent, like an alpha male, or an abstract goal such
as a task, a leadership role, such as a head of department, or
even a group mission statement, i.e. something which represents the abstract semantics of the group
itself.  Promise Theory therefore predicts
that repetitive interactions---what we call `grooming' of the
relationships---represent the exchange of information, learned by
participating agents (scaling to one to $N-1$ in a group of size $N$),
as reflected in section \ref{accounting}), leading to a group attachment
through the promise of mutual convergence for as long as that promise
can be considered trustworthy.  Once agents are connected through a
task or leader, they experience one another directly or indirectly.
Eventually, the cost of policing that contention from variational
noise between in the group over a broad bath of environmental
pressures can overwhelm individuals promise to accept the alignment
with the seed and the individual drifts away from the group.

\section{Cost accounting of `grooming' and contention}

Our interpretation of trust works as
an attention accounting quantity, driven by work done at different
times, past and present. It offers a complementary view to temporal
activity which is suitable for a statistical average independent of
time.  Trustworthiness (process potential energy) is a summary of
historically accumulated work of alignment with past promises,
expressed as a coarse snapshot of the slowly-varying history. It has
the semantics of reliability or believability for each individual
agent to assess. Conversely, attentiveness (process kinetic energy) is
an immediate release of work at some rate or velocity, in response to
residual uncertainty, expressed directionally by the potential alignment. 
The kinetic attention processes are
not linear velocity of motion in space but rather something associated with
cyclic Shannon information sampling \cite{shannon1}---a control
loop checking `are we there yet?'.

\subsection{Dimensional argument for the scaling of trust}

Quantitatively, the rate of conversion of accumulated work from the
keeping of promises must be found dimensionally, by comparing
the orders of magnitude, with the same engineering dimensions:
\beq
V &\sim& \2 mv^2\nonumber\\
v &\sim& \sqrt{V}
\eeq
up to dimensionless factors, where the dimensions of the
attention rate or `velocity' $v$ are arbitrary except for the role of time.
The potential amounts to a reliability for promise keeping, which we expect
to grow like the square root of the assessed trustworthiness up to some maximum sustainable size.

The work of a single agent, interacting in a group of size $N$, would
be expected to scale as \beq W\text{(agent)} =
\frac{c_1+c_2(N-1)+\ldots}{c_0 N_\beta} \eeq where $c_0$, $c_1$, and
$c_2$ are constants, and $N_\beta$ is some constant with the same dimension
as $N$. Our task is to determine these constant scales.  At low utilization, we can expect the
availability or channel capacity of the systems, viewed as an information
process, to be approximately proportional to
the number of agents interacting. Once contention sets in, this
capacity is depleted and the effective capture rate slows down.
Agents begin to leave a group, on average, leading to an equilibrium,
which is the value at which contention is maximal.  Although it is not
completely clear from the data that an order $(N-1)^2$ interaction is
excluded, the data fit with $(N-1)$ is less noisy for the numbers where we
have data. We take this as evidence that the dominant effect is from a
seeded process (see figure \ref{coop}(a)), and thus we neglect
additional parameters which contribute to noise.

When $\beta E = 0$, the probability has to be 1, so for $N=1$ (self), all the share is
in one agent's hands. So $c_1=0$. Now we have a single scale $C\equiv c_2/c_0$ representing
the level of shared of contention between agents. To determine this, we use the promise
seed configuration again below. Note that, at maximum entropy, a group is evenly distributed
by definition, 
without favour to any particular agent, so based on these dimensional arguments, 
we expect the large $N$ limit to take the form of a Boltzmann distribution:
\beq
P(\beta) \sim \exp\left(-\frac{\chi(N-1)}{N_\beta}\right),
\eeq
where we now see that the role of $N_\beta$ is that of a scale, which characterizes the intra-group contention. Small $\chi$ 
implies tolerance of contention, or loose coupling and thus larger group sizes (exactly as has been described for primate and human social groupings \cite{dunbar18,dunbar17}),
while large $\chi$ implies some kind of territorial overlap that leads to altercation.

\subsection{Work afforded by a limited capacity agent}\label{accounting}

Let us pursue the cost argument in terms of the physics of information.
Suppose each agent has a cognitive processing work capacity $W_\text{max}$
for the process of group interactions that it shares with other tasks too. How the capacity is sliced
is a detail that we don't need to address here, but if we think once again 
in terms of the energy analogy, about the `power output' or work done (i.e. the cost expended) 
by the agents to attend to one another, then the `kinetic' or spending rate terms can be related
to the promise of sharing the group resources in the following straightforward way. 

We assume that at large $N$ behaviour, averaged over large ensembles,
the probable work fraction $P(W)$ for distribution takes the form of a Boltzmann 
distribution over the relative costs\cite{reif1,treatise1}. We can write this in the form
familiar from physics texts:
\beq
P &\sim& e^{-\beta E}\\
\text{where (dimensionless)} ~~~ \beta E &\mapsto& \frac{W(\text{agent})}{\text{Total capacity for work}},
\eeq
though the key point is that it is a negative linear exponential.
Here the capacity or availability for expending work attention represents some finite budget for shared resource channel capacity.
Here, the dimensionless exponent is written traditionally as $\beta E$, from its thermodynamic origins
with $\beta$ as an inverse temperature ($\beta \sim 1/kT$, also called the coldness or thermodynamic beta) and $E$ as an energy. 
For us, these roles are used mainly for familiarity, in keeping with other literature.
We recall, from Shannon, that the channel capacity is a dimensionless representation 
of the channel's `power'\cite{shannon1}:
\beq
C = B \log \left( 1 + \frac{W(\text{agent})}{\text{Cost of contention}}\right)
\eeq
where $B$ is the maximum bandwidth for throughput, which is consistent with our assumption. 
With these points in mind, and assuming that interactions between group members are not `all at once', 
but interleaved approximately one at a time,
the accumulated work should be proportional to the group remainder size:
\beq
W_n \le \frac{W_\text{max}}{N},
\eeq
The bulk of this work is assumed to be the handling of contentious impositions\cite{promisebook} by group members
to reverse efforts and otherwise interfere with the agent's own alignment, either preventing or smoothing over such incidents.
The agent may have other things to deal with in addition to this `grooming' or placating of contentious others, so this work
allocation might not be 100\% efficient.
So we can take the cognitive capacity as a share for work:
\beq
(N-1)W_N = \2m v^2,
\eeq
for some rate $v$. Now, we arrange to measure these quantities in units such that we can
compare dimensionless ratios. In dimensionless form, we can compare the only matching scales
in the problem:
\beq
(N-1)\frac{W_N}{W_\text{max}} = \2 \frac{m}{m_\text{min}} \left(\frac{v}{v_\text{max}}\right)^2,
\eeq
The effective mass of the interaction (which plays the role of the 
cost of agent `involvement' with others) presumably has a minimum scale rather than a maximum,
though this doesn't matter since we eliminate this by changing variables.
None of these work rates are measurable in this study, so we need to relate them to something
with dimensions of $N$. We can make the identification
\beq
\frac{W_N}{W_\text{max}} \frac{m_\text{min}}{m} \equiv \frac{\beta}{\Nav}, \label{xx}
\eeq
which has the form
\beq
\frac{\text{Fractional work effort}}{\text{Fractional cost of involvement}}\times \text{efficiency},
\eeq
where we use the constant $\beta \le 1$ as an efficiency.
This is motivated by the identification of $\Nav$ as the scale for group size with
maximal contention cost.
From (\ref{xx}) we interpret the Dunbar group size as being based on:
\beq
\Nav = \text{cost as a fraction of work budget} \times \text{cognitive efficiency}.
\eeq
In the Wikipedia study in \cite{burgessdunbar1}, 
$\Nav$ is associated with data called the group contention cost, which is an emergent
scaling limit determined by looking to the group size at which contention arises, or when
$\Nav$ agents are all watching closely.
The actual value of the scale $\Nav/\beta$ has some arbitrariness between $\Nav$ and $\beta$,
so it can't be derived without a specific implementation model, but
we expect this is an innate internal capacity of each kind of agent, as originally
proposed by Dunbar. The universality of the expression, and the split can be seen in figure \ref{table}.

For example, agents may come together around a particular seed when their
prioritization of the seed promise becomes the dominant force in their
behaviour. Perhaps the appearance of a predator activates a
behaviour for a herd, or the appearance of a new Wiki page on a subject
close to one's heart activates a desire to contribute.  In the absence
of an attraction, there are enough alternative attractions to pull
animals away, leading to an exponential decay of this heightened
priority, typical of maximum entropy processes.

\subsection{Probability of occurrence for group size $N$}

We can now extend this argument to predict the
dimensionless frequency (or probability) of finding a group of size
$N$, by combining growth and decay, according to the equilibrium of the attachment and detachment rates. 
We denote this distribution by $\psi(N)$, or $\psi(\nu)$ for a dimensionless
variable $\nu$, defined below. This is the effective static
representation of the dynamic microscopic process of attachments represented by
a variable $\nu(N)$.
The result is a form of gamma
distribution.  We don't need to fit this by regression; we have
predicted the form necessary to comply with the scaling
of mutually aligned cognitive processes formed by seeding an arbitrary
direction. It no longer matters what the seed was, or who is aligning
with it. The probability refers only to a process of probabilistic
attachment and detachment, shaped by a trust potential which is learned
by a past history memory process. Growth is initially by invested kinetic effort in aligned order $\sqrt{\nu}$
and decay is by average disordered contention $\exp({-\nu})$.

\begin{figure}[ht]
\begin{center}
\includegraphics[width=12cm]{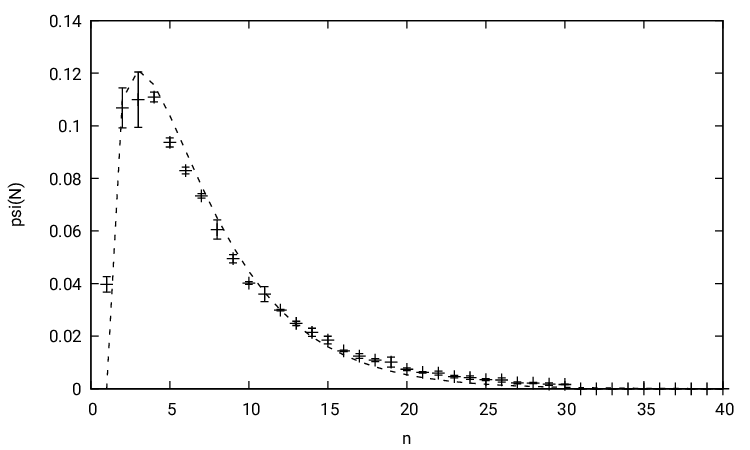}
\caption{\small Curve fit of data using the formula in equation \ref{formula} with data
from around 200,000 agents on Wikipedia. The crosses
approximate error uncertainty. The model fit is expected to be worst for small $N$ due to 
integer effects. In the Wikipedia results, $\beta = 1$ gives the appropriate fit.
In Dunbar's human groups. $\beta=0.75$ is a closer estimate of the promise efficiency.
\label{datafit}}
\end{center}
\end{figure}

The graph in figure \ref{datafit} fits very closely a simple formula,
in dimensionless form, which we can motivate from the theory:
\beq
\psi(\nu) = \frac{4}{\sqrt\pi}\; \frac{\nu^{\2}\; e^{-\nu}} {\langle N\rangle_{\overline T}}, 
~~~~~~\nu = \frac{2\beta(N-1)} {\langle N\rangle_{\overline T}} ~,~ (N > 1),\label{formula}
\eeq
where $\beta$ corresponds to a dimensionless (probabilistic) rate of promise keeping for the seed promise,
i.e. $\beta$ is the fraction of promises kept reliably, since reducing $\beta$ has the same effect as
reducing the group contention size limit (less tolerance of contention). Another way to think of $\beta$
is therefore as an metaphorical `coldness' to agent entropy, with effective energy
parameter $E = 2(N-1)/\langle N\rangle_{\overline T}$ for a group of $N$ agents. As contention increases, the
maximum occurs at smaller groups and that is equivalent to less effective promise keeping to interact with
the seed agent.
The result of this fit is shown in figure \ref{datafit}.

The relationship between the maximum frequency and maximal contention scales is determined
by the rate equation for detailed balance that leads to (\ref{formula}).
The value of $N$, which maximizes kinetic mistrust, is called $\langle N\rangle_{\overline T}$, while
the value of $N$ leading to the maximum value of $\psi(N)$, determined by $\frac{d\psi(N)}{dN}=0$ is:
\beq
N_i^\text{max} = 1 + \frac{\langle N\rangle_{\overline T}}{4\beta}.\label{maxn}
\eeq
Notice how the expected group size is still always less than the maximal contention size,
and that the group size, which maximizes productivity, in figure \ref{datafit}, is $N=4$.
There is a tradeoff between having a larger group and the cost of managing conflict.
This is interesting, as it suggests that (statistically) agents will
effectively tend to prioritize working more intimately with smaller
groups--like so so called pizza teams referred to in technology companies. 
This could be a sign that there is an additional contention
cost associated with switching between on going relationships, as
there is in computing---called {\em context switching}.

\subsection{The scaling of group hierarchy}\label{maxcurve}

We can examine some values for these maxima relationships 
to illustrate the fit with the layer model in Dunbar\cite{dunbar3}
and the specific data for Wikipedia\cite{burgessdunbar1}. The column for $\beta = 1$ reproduces the results from
the Wikipedia data in \cite{burgessdunbar1}. 
Removing all non-human 
automated software robots or `bot' interactions alters $\Nav$ slightly to give an effective value of $\beta=0.93$.
The column with lower efficiency $\beta=0.875$ generates the usual stylized Dunbar sequence quite accurately (see table \ref{table}).

\begin{table}[ht]
\begin{center}
\small
\begin{tabular}{c|c|c|c}
Mode & Bots+Humans & No Bots & Humans (Dunbar) \\
\hline
$N_i^\text{max}$ & $\langle N\rangle_{\overline T}~ (\beta=1)$& $\langle N\rangle_{\overline T}~ (\beta=0.93)$ & $\langle N\rangle_{\overline T}~ (\beta=0.875)$ \\
\hline
3 & 8 & &\\
5 & & 14.9 & 14\\
8 & 28 & &\\
14 & & & 45.5\\
14.9 & & 52 &\\
28 & 108 &&\\
45.5 & && 156\\
52 & & 188 &\\
108 & 428 &&\\
156 & & & 542\\
188 & & 697 &\\
428 & 1708 &&\\
542 && &1892\\
\end{tabular}
\caption{\small A table of predicted hierarchy of group statistical size and 
contention maxima for three different parameters involving human interactions. See figure \ref{hierarchy}
for a schematic of how to read the sequences. Each parameter column
makes a close fit with data selections with different average cognitive capacities. The figures
for purely human interactions match the numbers for the Dunbar hierarchy most closely, and those
involving bots with artificial cognitive stamina indicate a slightly higher tolerance for average group size.
\label{table}}
\end{center}
\end{table}

\begin{figure}[ht]
\begin{center}
\includegraphics[width=5cm]{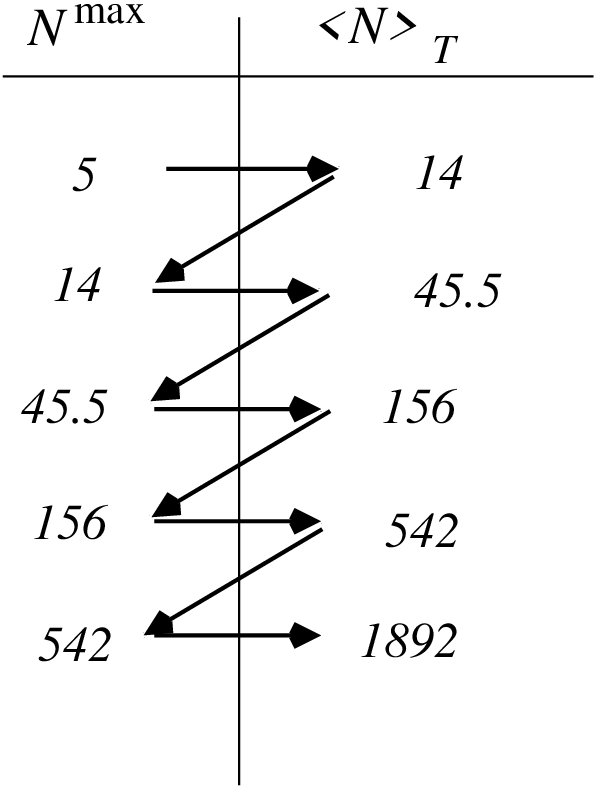}
\caption{\small A schematic showing how to read the hierarchy in figure \ref{table}.\label{hierarchy}}
\end{center}
\end{figure}

\begin{figure}[ht]
\begin{center}
\includegraphics[width=11cm]{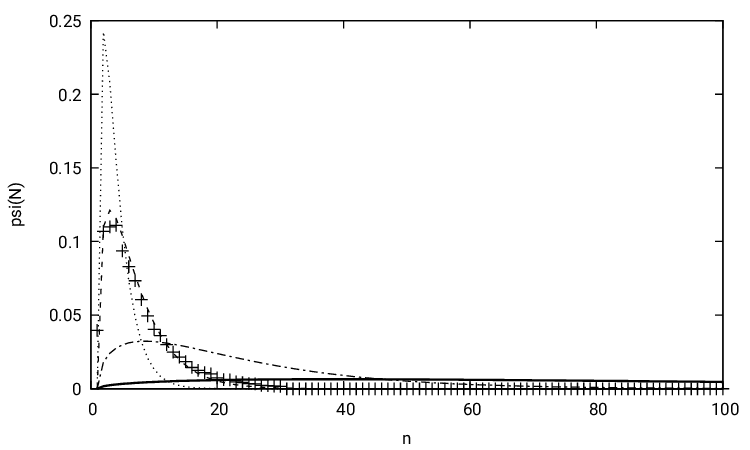}
\caption{\small The group equilibrium law plotted for $\langle N\rangle_{\overline T} = 4,8,30,150$
illustrating the flattening of group probability curves with increasing number. The amplitude
gives an approximate magnitude for the attention power rate required to maintain each level.\label{dunbarall}}
\end{center}
\end{figure}

Reading down each column, we see the mode frequency limited by the
next scale up in the two right hand columns.  We note that the
apparent self-similar scaling fraction of group sizes depends on
$\beta$ for its precise value in equation \ref{maxn}. The specific
work discussion related values from Wikipedia are slightly above the
multi-case average values summarized by Dunbar\cite{dunbar6a}, but are close to the
more specific results for conversations \cite{dunbar7a}. See also the schematic
scale transformation series in figure \ref{hierarchy}, and combined plot in figure \ref{dunbarall}.

\section{Trust, attention, and neural processes}

By calling the work of attention a kinetic process, readers might get
the false impression that it implies the effort of agents running
around or performing manual labour. However, we recall the original
Dunbar hypothesis that the relevant work processor is the amount of primate neo-cortical
mass.  If group size is moderated by a process of contention between
agents and, implicitly, grooming is work invested in overcoming it
(as has been shown to be the case in both primates and humans, where friction and homicide rates increase linearly with group size \cite{dunbar26,dunbar27}),
then it certainly isn't the work of picking nits out of fur that
accounts for the group scaling. It can only be the work of memorizing
the identities and foibles of the individuals in an environment: the
cyclic process of building trust from distinguishing individuals from a faceless background.  This
prompts a natural speculation that arises from our quantitative
prediction: we are led to ask what are the dominant neural processes
at each level of the hierarchy? One possibility could be that the
group sizes correspond to different level of brain activity.  Moreover, a natural proxy
for these dynamics is perhaps the cyclic `brainwave' oscillation modes\cite{buz0}
for the transport of information between cortical regions (see figure
\ref{neuro}).

Frequency is associated with power\cite{motokawa}, so it's interesting to 
compare the hierarchy of group sizes to the power associated with levels of
attention or brain concentration. Buzs\'aki writes \cite{buz1}:
``The power density of local electrical field potential is inversely
proportional to frequency in the mammalian cortex.  This 1/f power
relationship implies that perturbations occurring at slow frequencies
can cause a cascade of energy dissipation at higher frequencies and
that widespread slow oscillations modulate faster local event.'' Thus
the idling work required for attentiveness in a typical group size
might be expected to follow the same kind of power
requirement.
Once again, on dimensional grounds $\Nav$ can only appear in this
relationship multiplied by an effective time conversion scale $\Delta\tau$
for the `latency', and the product of this with frequency  $f\times\langle N\rangle$ represents
an average throughput of information up to some intrinsic timescale $\Delta\tau$. So in relative units
we get the results in figure \ref{neuro}.

\begin{figure}[ht]
\begin{center}
\begin{tabular}{c|c|c|c}
Attention & Brain wave (Hz) $f$ & Dunbar $\langle N\rangle$ level & $f\times\langle N\rangle$\\
\hline
light attention& $\alpha$  5-15 (5)   & 150  & 750 \\
middle attention &$\beta$  12-30 (25)   & 30 & 750\\
 concentrated & $\gamma$-fast 32-200 (150)  & 5& 750\\
\end{tabular}
\end{center}
\caption{\small Approximate brain process oscillation frequencies associated with heightened attention.\label{neuro}}
\end{figure}
As a rough guide, we can calculate an optimistically tendencious product of the columns
and the result is indeed of approximately of constant order, as long as we
cherry pick from the somewhat broad ranges. A more careful study would require
expertise we don't have ourselves, but this is at least suggestive that the
average effort is indeed in inverse proportion to the group size. This
is numerically interesting, if not exactly proof of a connection.

\section{Conclusions}

What began as effort to understand the meaning of trust in social
dynamics, through the lens of agent attention rates in Promise Theory, has led us
to an explanation for the hierarchy of social group sizes discovered
by Dunbar, and allows us to see how the mechanism applies beyond that
scope. In this work, we can see these two narratives as part of a 
conjoined phenomenon, thus offering a tantalizing perspective on each.

Our model makes a bold assumption, supported by the scaling, namely
that groups in a social brain hierarchy are not random processes, but
are formed around a seed of {\em intent}, which acts to capture the attention
of agents through associated kinetic processes.  There is thus a de facto
attractive `force' that promotes group accretion on a small scale, and
later fades away to become asymptotically free as groups disband.
Agents offer their attention to group processes variably in order to
invoke a simple optimization for beneficial reasons. They have a
finite budget for attention, which is governed by their work capacity.
Our model shows how we can relate microscopic and macroscopic pictures
for one class of behaviours, as a kind of kinetic theory or
statistical mechanics of social interactions. On a small scale, social
groups come together in response to an initial seed that attracts the
attention of agents. The group accretes new members until contention
between them eventually drives the group apart or the seed loses its
interest value.  In particular, we calculate the probability for
reaching a certain group size, based on the work expended in attending
to other agents.

What is the all important seed promise for a group? By definition, it
promises the role of a prioritized behaviour that's shared by the
individuals in a group.   It only takes one agent to start an activity for others
to follow. Then groups grow by accretion.  How do agents come
together?  In the case of Wiki editing, it's clearly the promise of
the platform to enable satisfactory publishing of information---the
creative commons, with its attendant benefits.  For animals in a pack
or herd, it might be the promise of a defensive posture when a
predator is nearby, or the co-location of some tidbit, that drives
them to attend to one another's relative positions and cluster. They
would then drift apart again once the seed were gone\cite{jackdaw}.
For a religious group or company, it could be a charismatic
leader\cite{dunbar4}, which also aligns with work on the origin and
semantics of authority\cite{burgessauthority1}.  Alternatively, it
could be a more abstract health benefit acquired as an evolutionary
adaptation over very long times, such as when a change in the weather
or other environmental conditions triggers group changes, as in slime
mould dissociation for instance---or merely the opportunistic sharing
of a transient resource\cite{ostrom1}.  The semantics of a seed of
intent might change frequently to reflect changing group dynamics,
even as the underlying dynamics remains a universal function of
physiology.  These are cases where Promise Theory's agenda of unifying
dynamics with semantics seems particularly well suited.

In a future in which humans bond with artificial enhancements as
`cyborgs', Artificial Intelligence may alter some aspects of the scales here.
This could, in turn, pose a different spectrum of challenges to human
character that needs exactly the kind of cognitive capacity predicted
in the Dunbar hierarchy to deal with effectively.
Given that we have shown that relationship between cognitive effort
and group size is not strongly dependent on an affinity to any specific
species or details, we can speculate about the implications of the model
in such other cases. What, for instance, would the same limits
mean for the execution of other mentally taxing tasks, particularly where
groups and teams are concerned? How are the limits affected by the 
introduction of artificial reasoning and automation? In \cite{burgessdunbar1},
we saw automation inflate group sizes slightly, but as long as humans were involved
human limits were in play.

Clearly attachment is not the only mechanism for group size either. Kin are
formed by `budding' rather than by accretion, which creates rather special
semantic bond. Yet families are not immune to dispersal, particularly
where impersonal technologies are able to overwhelm 
individual communication. Broadcast media and modern network media
channels clearly have the capability to change the dynamics of
societal cohesion. There might thus be
processes that can overwhelm the specific formula we derive here, but they
currently lie in wait to take over in changed circumstances.

Our results offer an objectified causal explanation for the
empirically demonstrated fact that human communities have multiscale
`fractal' levels.  Moreover, they decouple and `de-personalize' the
issues in such a way that the model manifestly applies to other kinds
of agency, with other kinds of cognitive processing--such as social
institutions, cities, and cybernetic systems.  The phenomenon arises
because (a) there are constraints on the time available for individual
agents to interact, (b) there is a strong preference for distributing
what time we have unequally among potential alters in ways that
reflect the benefits we expect to obtain from them in the
future\cite{dunbar22}, and finally, (c) in addition in sentient agents
the willingness to spend time on relationships with others is strongly
dictated by our emotional `warmth' towards them\cite{dunbar23}.  There
are clearly implications here for understanding the opportunities for
innovation and cultural exchange, as well as political affinities, and
so on.  

\changes{In the end, one way of stating the conclusion of this model
  is that a social group is a form of tool for economic benefit, and
  the other tools we create build on the cognitive underpinnings of
  social interactions.  Indeed, in the modern world, we often spend
  more time working with and getting to know our tools and processes
  than with friends and family members. One important implication of
  our findings from the Wikipedia editors study is that work groups
  (sets of individuals collaborating on a task) are strictly limited
  in size to around four individuals. Larger groups fail to coordinate
  effectively, are more prone to disagreements and conflicts and
  consequently shed members rather than recruit new ones. This finding
  has profound implications for how we organize work groups in order
  to maximize production of technology.  Earlier hominin species had
  much smaller brains than modern humans, as Koppl et al.
  \cite{koppl} noted, implying that that they formed much smaller
  groups \cite{dunbar24} and similarly had less capacity for in-depth
  working relationships, limiting both the rate of novel innovations
  and the rate at which these would have diffused through the wider
  population\cite{dunbar25}.}

Ultimately, we might note that progress in theoretical social science
has been slow compared to that of the other natural sciences. As a
nascent field, socio-physics\cite{galam,sociophysics2} shows some
successes; however, socio-physics argues principally by fortuitous
analogy to known phenomena in physics. Here, we have been able to
provide a missing piece of explanation for a genuine scientific model,
with an underlying causal justification for such similarities---they
are no longer merely fortuitous.  Our universality argument, built on
the Promise Theory of agent interactions with trust as a currency of
economic accounting, provides just that missing link.

\bigskip

Acknowledgment: MB is grateful to Gy\"orgy Busz\'aki for discussions about the neuroscience.
This work was supported by a grant from the NLnet Foundation (2023).
RD acknowledges funding by ERC Advanced Research grant (\#295663).

\bibliographystyle{unsrt}
\bibliography{bib1,bib2}

\end{document}